\documentclass[preprint]{aastex}
\newcommand{\objD}{J0134--0931}

\slugcomment{}

\begin{document}

\title{ PMN~\objD: A gravitationally lensed quasar\\with an unusual radio
morphology }

\author{
Joshua N.\ Winn\altaffilmark{1,2},
James E.J.\ Lovell\altaffilmark{3},
Hsiao-Wen Chen\altaffilmark{4},
Andr\'{e} B.\ Fletcher\altaffilmark{5},
Jacqueline N.\ Hewitt\altaffilmark{1},
Alok R.\ Patnaik\altaffilmark{6},
Paul L.\ Schechter\altaffilmark{1,2}
}

\altaffiltext{1}{Department of Physics, Massachusetts Institute of Technology,
    77 Massachusetts Avenue, Cambridge, MA 02139, USA}
\altaffiltext{2}{Visiting Astronomer, Cerro Tololo Inter-American Observatory,
    National Optical Astronomy Observatories, Casilla 603, La Serena, Chile}
\altaffiltext{3}{Australia Telescope National Facility, CSIRO, PO Box 76,
    Epping, NSW 1710, Australia}
\altaffiltext{4}{The Observatories of the Carnegie Institution of Washington,
    813 Santa Barbara St., Pasadena, CA 91101, USA}
\altaffiltext{5}{MIT Haystack Observatory, Off Route 40, Westford, MA 01886,
    USA}
\altaffiltext{6}{Max-Planck-Institut f\"{u}r Radioastronomie,
    Auf dem H\"{u}gel 69, 53121 Bonn, Germany}

\begin{abstract}
The radio-loud quasar \objD~was discovered to have an unusual
morphology during our search for gravitational lenses. In VLA and
MERLIN images, there are 5 compact components with maximum separation
681~mas. All of these components have the same spectral index from
5~GHz to 43~GHz. In a VLBA image at 1.7~GHz, a curved arc of extended
emission joins two of the components in a manner suggestive of
gravitational lensing. At least two of the radio components have
near-infrared counterparts. We argue that this evidence implies that
\objD~is a gravitational lens, although we have not been able to
devise a plausible model for the foreground gravitational
potential. Like several other radio-loud lenses, the background source
has an extraordinarily red optical counterpart.
\end{abstract}


\section{Introduction}
\label{sec:intro}

Gravitationally lensed quasars are rare and valuable astrophysical
tools. They can be used to determine the Hubble constant
(\citealt{refsdal64}) and place limits on the cosmological constant
\citep{turner90,fukugita90}. The image configuration reveals
properties of the matter distribution of the foreground galaxy,
including dark matter, and in some cases gravitational magnification
reveals interesting details that would otherwise be too faint to
observe (for reviews see, e.g.,
\citealt{blandford92,wambsganss98,narayan99}).

Lenses are usually recognized as such by an unusual morphology in a
high-resolution ($< 1\arcsec$) image. A quasar that exhibits two or
more components in an optical or infrared image is an excellent lens
candidate. Likewise, a radio source with more than one compact
flat-spectrum component, or a radio lobe with a ringlike or arclike
morphology, is likely to be the result of gravitational lensing.

Such candidates are usually followed up with imaging at multiple
wavelengths and/or higher resolution, and with optical
spectroscopy. The goals are to determine whether the components have
identical spectra and surface brightness, as would be the case for
lensed images, and to search for evidence of foreground objects that
are responsible for the gravitational deflection.

This strategy has been employed by many lens surveys, including our
own, which is a radio-based survey of the region $0\arcdeg > \delta >
-40\arcdeg$. We used the VLA\footnote {The Very Large Array (VLA) and
Very Long Baseline Array (VLBA) are operated by the National Radio
Astronomy Observatory (NRAO), a facility of the National Science
Foundation (NSF) operated under cooperative agreement by Associated
Universities, Inc.} to search for characteristic lensing morphologies
in a sample of approximately 4000 radio sources, chosen to be
flat-spectrum between the 4.85~GHz PMN catalog \citep{pmn} and the
1.4~GHz NVSS catalog \citep{nvss}. This paper reports the discovery of
PMN~\objD, the fourth object in our survey that we believe is
gravitationally lensed; the others are J1838--3427 \citep{j1838},
J2004--1349 \citep{j2004}, and J1632--0033 \citep{j1632}.

Independently, \objD~was identified by \citet{gregg01} during a search
for highly reddened quasars. In a companion paper, they present
optical and near-infrared spectra exhibiting quasar emission lines at
$z=2.216$, and a $K'$-band image revealing at least two components.
These authors also argue that the quasar is being gravitationally
magnified, and examine the possible implications of the extreme
redness of \objD.

The next section describes the radio properties of \objD.
Section~\ref{sec:opt} describes the near-infrared and optical
properties. Section~\ref{sec:hypothesis} argues the case for
gravitational lensing based on this evidence and points out
complications of the lensing hypothesis. Section~\ref{sec:summary}
contains a summary and suggests future observations to test the
lensing hypothesis and constrain lens models.

\section{Radio properties}
\label{sec:radio}

\subsection{VLA, MERLIN, and ATCA observations}
\label{subsec:vla}

\begin{figure}
\figurenum{1}
\plotone{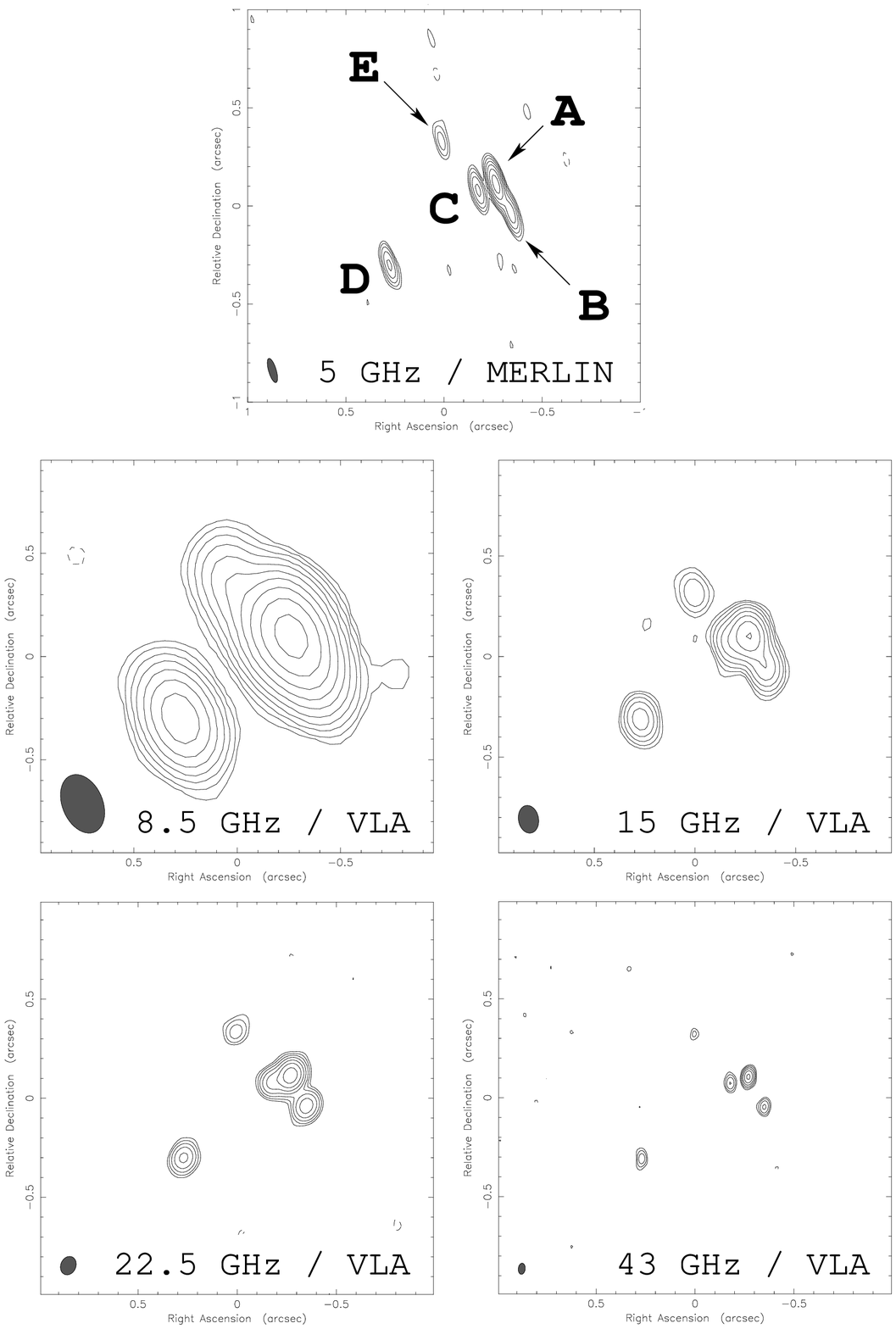}
\vskip -0.5in
\caption{ Radio contour plots of \objD~with the VLA and MERLIN.  In
all cases, the field of view is $2\arcsec \times 2\arcsec$, the image
is based on uniform weighting, and the synthesized beam is inset in
the lower left of each panel. Contours begin at $3\sigma$ and increase
by factors of 2, where $\sigma$ is the RMS noise level (see
Table~\ref{tbl:radio}). }
\label{fig:radio}
\end{figure}

We selected \objD~as a lens candidate because of its unusual radio
morphology. Table~\ref{tbl:radio} provides a summary of our radio
observations, and also lists the coordinates of
\objD. Figure~\ref{fig:radio} presents radio images with the VLA and
MERLIN\footnote{ The Multi-Element Radio Linked Interferometry Network
(MERLIN) is a UK national facility operated by the University of
Manchester on behalf of SERC.}, at frequencies ranging from 5~GHz to
43~GHz.  The details of calibration and data reduction were as
follows:

For the VLA and MERLIN observations, we interleaved observations of
\objD~with the nearby source J0141--0928 in order to calibrate the
antenna gains. To calibrate the absolute flux density scale for the
MERLIN data, we observed 3C286 and assumed a flux density of 7.38~Jy
on the shortest baseline. For the VLA observations in October 2000, we
observed J2355+4950 instead; this secondary flux calibrator is
monitored monthly by G.\ Taylor and S.\ Myers of NRAO, and has been
found to have a stable flux density. We assumed the flux densities of
this source (in Janskys) were 0.921, 0.602, 0.473, and 0.284 at
8.5~GHz, 15~GHz, 22.5~GHz and 43~GHz, respectively. Calibration was
performed using standard procedures within the software package
AIPS. We applied gain-elevation corrections for data at 15~GHz and
higher frequencies, based on gain curves prepared by S.\ Myers.

Imaging was performed with the software package Difmap \citep{difmap}.
Five radio components were apparent in each image, except the 8.5~GHz
image, in which four of the components were blended together. In
Figure~\ref{fig:radio} these components are labelled A--E, in
decreasing order of flux density.  This order is the same in all five
images; indeed, the visual appearance of the images suggest that all
the components of \objD~have approximately the same radio ``color.''

To investigate this point quantitatively, we fit a model consisting of
5 point components to the visibility function of each data set. We
used this model to perform phase-only self-calibration with a solution
interval of 30 seconds. This process, model-fitting and
self-calibration, was repeated (typically 5--10 times) until the model
converged. In all cases, the relative separations of the 5 components
(printed in Table~\ref{tbl:radiomodels}) agreed within 2~mas.

\begin{figure}
\figurenum{2}
\vskip -0.75in
\plotone{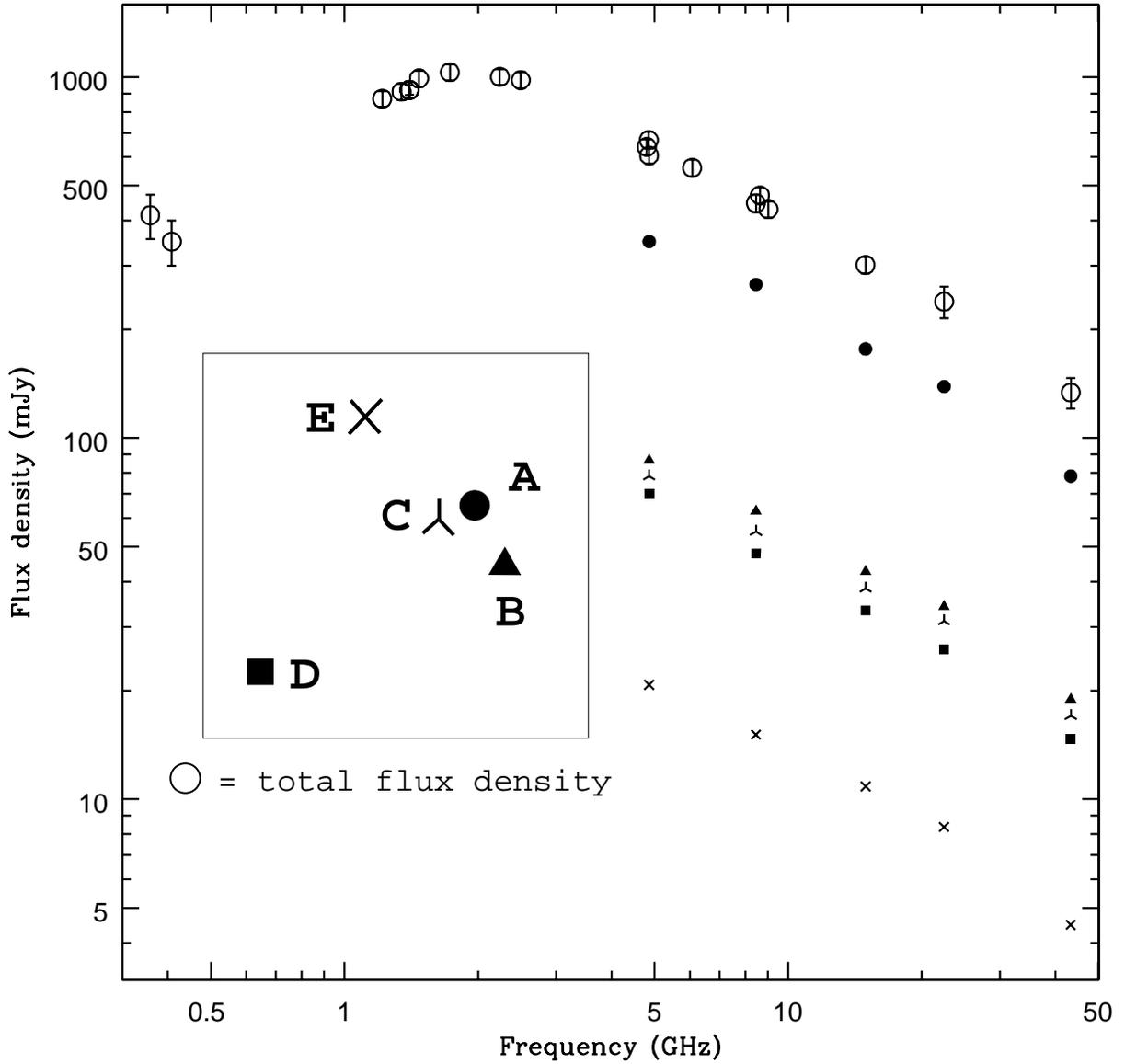}
\caption{ Logarithmic plot of the total flux density of \objD~(open
circles) and each of the 5 components A--E, as a function of radio
frequency. The legend (inset) is a model based on the VLA data. Error
bars shown for total flux measurements represent the uncertainty in
flux calibration, which affects all components equally. Where error
bars are not shown they are comparable to or smaller than the symbol
size. Total flux measurements are drawn from our own measurements and
the Texas~365~MHz catalog \citep{texas}, the NVSS 1.4~GHz catalog
\citep{nvss}, the FIRST 1.4~GHz catalog \citep{first}, the
PMN~4.85~GHz tropical catalog \citep{pmnt}, and a 408~MHz measurement
from 1978 with the Molonglo cross (R.\ Hunstead, private
communication). }
\label{fig:fluxes}
\end{figure}

The flux density of each component as a function of radio frequency is
plotted in Figure~\ref{fig:fluxes}. We computed the best-fit power law
$S_\nu \propto \nu^{\alpha}$ for each component. They are all
consistent with $\alpha = -0.69\pm 0.04$ between 5~GHz and 43~GHz.
Evidently, the radio continuum spectra of all five components have the
same slope. This comparison can be made more precise by comparing the
flux density ratios between components at each frequency, because
these ratios are not affected by the uncertainty in the absolute flux
scales. The ratios relative to component D are printed in
Table~\ref{tbl:radiomodels}. The differences between the ratios
measured at different frequencies are less than 4\%.

Also plotted in Figure~\ref{fig:fluxes} is the total flux density of
\objD~over a wider frequency range. The data were drawn from the
literature, our VLA and MERLIN measurements described earlier, and our
ATCA\footnote{The Australia Telescope Compact Array (ATCA) is part of
the Australia Telescope which is funded by the Commonwealth of
Australia for operation as a National Facility managed by CSIRO.}
measurements of 2000~September~25. The ATCA measurements were
performed when the array was in the 6D configuration, and used
PKS~B1934--638 to set the flux density scale. The total flux density
measurements reveal that \objD~is a gigahertz-peaked-spectrum (GPS)
source with a peak at 2~GHz. The morphology of \objD~is unusual for a
GPS quasar, or indeed for any radio source. Although morphologies of
GPS quasars are diverse, they are predominantly compact sources,
core-jet sources, or linear triples with an angular extent of 10--100
milliarcseconds \citep[see, e.g.,][]{odea98,stanghellini97}.

Two other characteristics of typical GPS sources are a low level of
centimeter-wavelength polarization, and low or non-existent
variability of total flux density. The lack of a polarized signal in
the MERLIN observations implies that the fractional polarization of
the brightest component of \objD~is less than 3\%. (The VLA
observations were not calibrated for polarization.) The agreement of
many measurements of the total flux density spanning almost 20 years,
as shown in Figure~\ref{fig:fluxes}, suggests that \objD~is not
significantly variable.

To summarize the main conclusions that we have drawn from the VLA and
MERLIN data: (1) \objD~is a GPS radio source. (2) It consists of 5
compact components in an unusual triangular arrangement with maximum
separation 681 mas. (3) The spectral indices of the components are the
same from 5~GHz to 43~GHz.

\subsection{VLBA observations}
\label{subsec:vlba}

We observed \objD~with the VLBA on two separate occasions. On
24~April~2000, we observed for one hour at 5~GHz, using 8 antennas
(the Mauna Kea and Hancock antennas were unavailable). On
31~October~2000, we observed for four hours at 1.7~GHz with all 10
antennas. The key parameters of these observations are printed in
Table~\ref{tbl:radio}. In both cases, the total observing bandwidth of
64~MHz was divided into 8 intermediate frequency bands of width 8~MHz,
each of which was subdivided into 16 channels of width 500~kHz. The
sampling time was 1 second.

Calibration was performed with standard AIPS procedures.  We solved
for phase delays and rates using a fringe-fit interval of 2 minutes.
After fringe fitting, we averaged the data into frequency bins of
width 2~MHz, and time bins of width 6~seconds. These values were
chosen to reduce the data volume as much as possible while keeping the
amount of bandwidth smearing and time-average smearing below $1\%$
over the required field of view.

\begin{figure}
\figurenum{3}
\plotone{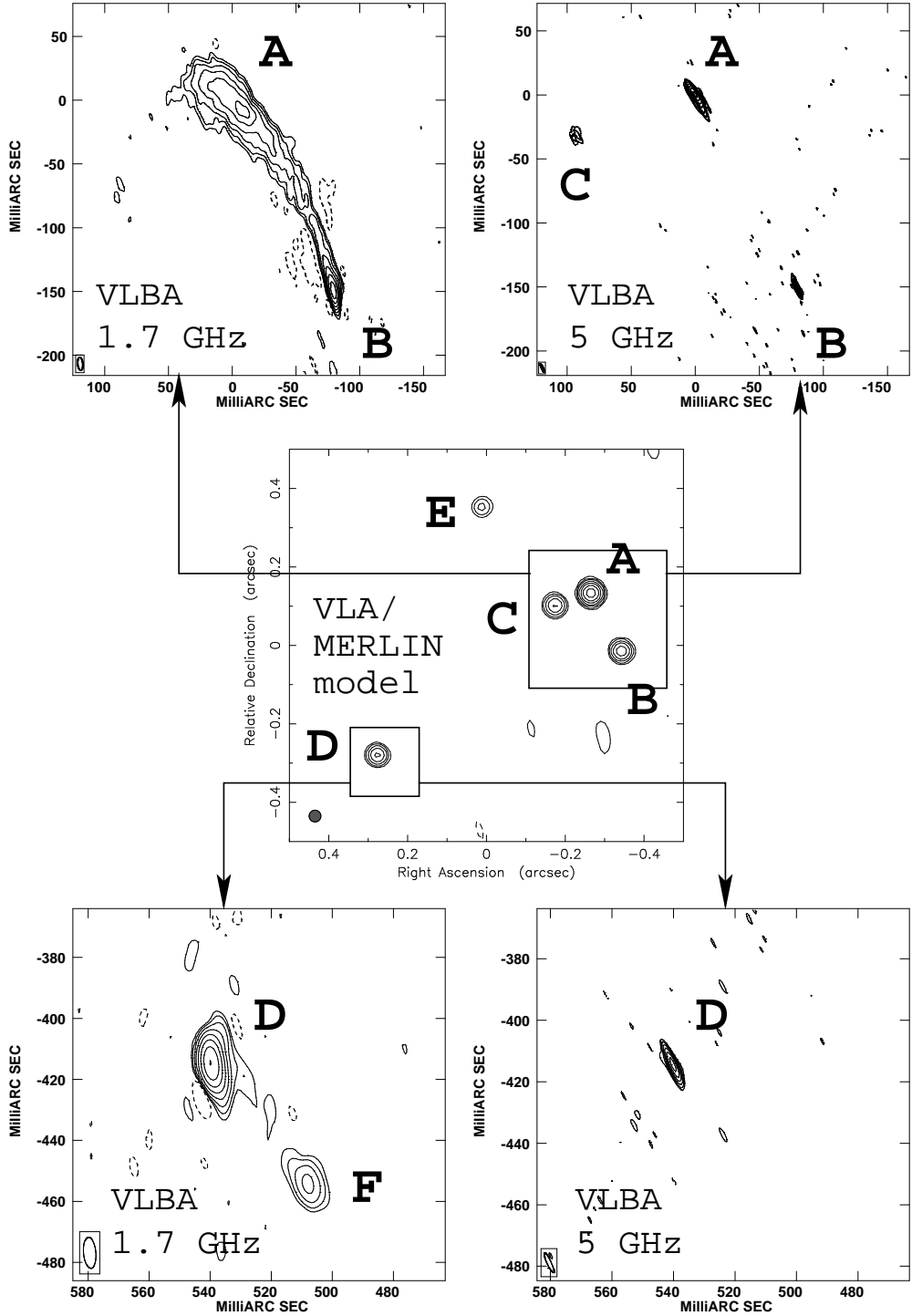}
\vskip -0.5in
\caption{ Radio contour plots of \objD~with the VLBA.  The center
panel is an illustration of the $2\arcsec\times 2\arcsec$ field based
on the VLA and MERLIN models.  The other four panels are enlargements
of the regions near components A,B,C (top) and D (bottom).  The
synthesized beam is inset in the lower left of each contour
plot. Contours begin at $3\sigma$ and increase by factors of 2, where
$\sigma$ is the RMS noise level. }
\label{fig:vlba}
\end{figure}

For imaging, we employed standard AIPS procedures. The process of
``cleaning'' (deconvolution) and phase-only self-calibration (with a
30-second solution interval) was iterated 3 times.
Figure~\ref{fig:vlba} presents the final images, using uniform
weighting. The central panel is not an image; it is merely an
illustration of the 5-component model developed from the VLA and
MERLIN data, to provide a wide-field overview of the system.

We first discuss the 5~GHz data. The four components A--D were
detected.  Components B and D are nearly unresolved, whereas component
A is highly elongated. Component C is barely visible in the image, and
has a larger angular size (and lower surface brightness) than the
other components. Table~\ref{tbl:vlba-6cm} lists the parameters of a
simple model consisting of 4 elliptical Gaussian components that fits
the data fairly well. In this model, component A was represented by 2
elliptical Gaussians. Component E is absent from the image, although
it does appear (faintly) when the data are re-weighted so as to
emphasize the shortest baselines.

In general, radio structure that is smooth on an angular scale greater
25--30 beamwidths is ``resolved out'' and will be invisible in VLBA
images. In order to gauge whether any of the components are
significantly resolved out, we divided the total 5~GHz flux of each
component as measured by the VLBA by its 5~GHz flux as measured by
MERLIN. The results are 1.00, 0.88, 0.50 and 0.95 for components A--D
respectively, confirming that C is largely resolved out. The
non-detection of E implies that its angular size must be larger than
about 35~mas. Because E was unresolved in the 5~GHz MERLIN image, the
angular size cannot be much larger than 80~mas.

\begin{figure}
\figurenum{4}
\plotone{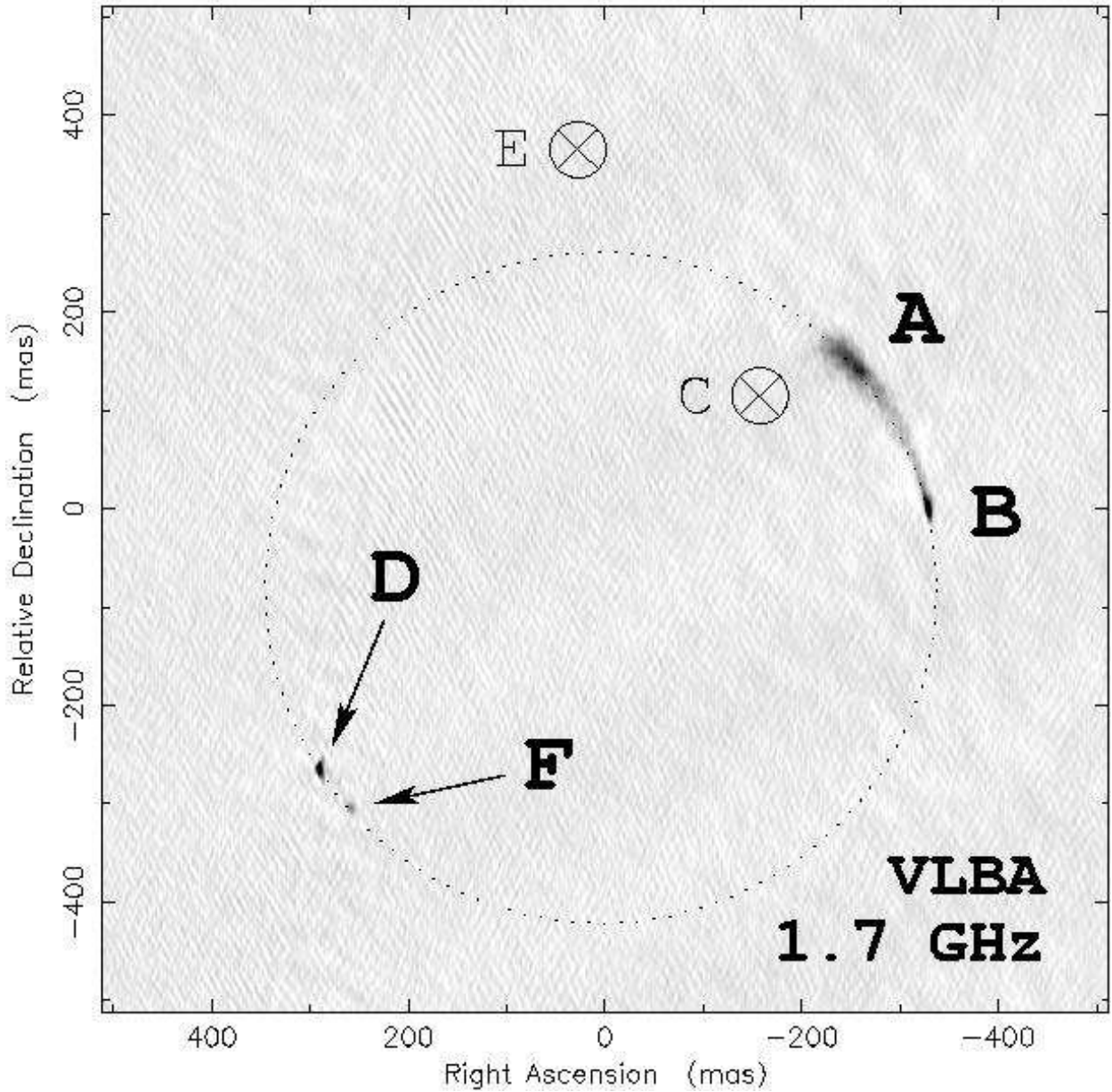}
% \vskip -1in
\caption{ Radio image of \objD~with the VLBA at 1.7~GHz.  Based on the
same data as the left panels of Figure~\ref{fig:vlba}, but here the
entire field is shown. The expected positions of components C and E
are marked. A dotted circle has been drawn to illustrate the
discussion of \S~\ref{subsec:vlba}. }
\label{fig:vlba-18cm}
\end{figure}

We next discuss the 1.7~GHz image. Figure~\ref{fig:vlba} contains
radio contour plots, and Figure~\ref{fig:vlba-18cm} is a wide-field
grayscale image. Components B and D are compact, but C and E are
completely absent. Component A is very extended and is connected to B
by a curved arc of emission.  Furthermore, a dim sixth component is
evident to the southwest of D, labelled F in Figures~\ref{fig:vlba}
and~\ref{fig:vlba-18cm}.

In Figure~\ref{fig:vlba-18cm}, a dotted circle has been drawn through
components A, B, D and F. This illustrates that all 4 components lie
on the same circle, and also that the arc joining A and B lies on the
circle. Both of these properties are strongly suggestive of
gravitational lensing (see \S~\ref{subsec:case}).

Table~\ref{tbl:vlba-18cm} lists the parameters of a simple model
consisting of 4 elliptical Gaussian components. This model provides a
decent fit to the image, but does not account for most of the flux in
the arc. The total flux density of the model is 0.634~Jy, which is
91\% of the total flux density in the image (0.696~Jy). Components D
and F are elongated towards one another. It is possible that D and F
are connected by an arc, as are components A and B, but that the arc
is too dim to be evident.

Judging from the total flux densities plotted in
Figure~\ref{fig:fluxes}, the total flux density of \objD~at this
frequency is about 1.0~Jy. The 0.3~Jy that is missing from the 1.7~GHz
VLBA image is mainly due to the absence of components C and E. These
two components would be expected to contribute about 0.2~Jy at
1.7~GHz, by extrapolating their flux densities measured at higher
frequencies. For these components to be resolved out by the VLBA,
their angular sizes must be about 120~mas or larger.

\section{Optical and infrared properties}
\label{sec:opt}

\subsection{Near-infrared counterparts}
\label{subsec:nearir}
We obtained an $H$-band image of \objD~on 2000~October~13 at Las
Campanas Observatory, using the Cambridge Infra-Red Survey Instrument
(CIRSI) camera mounted on the Cassegrain focus of the du~Pont
2.5-meter telescope.  Four exposures of 30 seconds were taken at each
of 5 dither positions, making for a total integration time of 10
minutes in the final stacked image. The pixel scale is $0\farcs200$
and the resolution in the stacked image is $0\farcs46$. A subraster of
the stacked image is shown in Figure~\ref{fig:ir}, along with a
contour representation.

The $H$-band counterpart is obviously elongated. There is also a dim
object $3\arcsec$ to the southwest, which is radio-silent in all our
radio images. We used the DAOPHOT routines in the software package
IRAF\footnote{ IRAF is distributed by the National Optical Astronomy
Observatories, which are operated by the Association of Universities
for Research in Astronomy, Inc., under cooperative agreement with the
National Science Foundation.  } to construct an empirical PSF of
diameter $10\arcsec$ using a star $50\arcsec$ east and $78\arcsec$
south of \objD. A contour representation is displayed in the lower
right panel of Figure~\ref{fig:ir}. We used this empirical PSF to
find the best-fit positions and fluxes of a model with three
components (two for \objD~and one for the dim object to the
southwest).

In this model, the double representing \objD~had separation $613\pm
75$~mas, position angle $126\arcdeg \pm 7\arcdeg$, and flux ratio
$6.3\pm 0.5$. The corresponding radio values for components A and D
are $681\pm 2$~mas, $127\fdg3 \pm 0\fdg2$ and $5.4\pm 0.1$. The other
pairings of radio components that agree with the $H$-band separation
are A/F, C/D, and C/F, but in those cases the radio flux ratios do not
match the $H$-band flux ratio at all.

The simplest conclusion is that A and D, at least, have near-IR
counterparts. It is possible that components B and C also have near-IR
counterparts that are merged with A in our image. Indeed,
\citet{gregg01} present an analysis of a $K'$-band image of \objD~that
supports this interpretation. After performing a maximum-entropy
deconvolution, these authors conclude that components A, B and D (at
least) have $K'$-band counterparts.

If \objD~is indeed gravitationally lensed, it is also possible that
the foreground galaxy contributes significantly to the total
light. These possibilities would complicate the interpretation of the
near-infrared images. For the purpose of evaluating possible lensing
scenarios, it will be important to obtain higher-resolution optical or
near-infrared images, using adaptive optics or the {\em Hubble Space
Telescope}.

To establish the zero-point for the $H$-band magnitude scale, we
measured the flux of star \#9106 described by \citet{persson98} within
an aperture of diameter $10\arcsec$. The resulting total $H$-band
magnitude of \objD~is given in Table~\ref{tbl:photometry} along with
the results from measurements through other filters.

\subsection{Optical counterpart}
\label{subsec:photometry}

On 2000~July~26, we obtained $BVRI$ images of \objD~with the Mosaic II
CCD~camera at the prime focus of the Blanco 4-meter telescope at
CTIO\footnote{The Cerro Tololo Inter-American Observatory (CTIO) is
operated by the Association of Universities for Research in Astronomy
Inc., under a cooperative agreement with the NSF as part of the
National Optical Astronomy Observatories.}. Each exposure lasted 10
minutes. The seeing varied from $0\farcs8$ to $0\farcs9$.  The images
were processed with standard IRAF routines, and the $I$-band image was
defringed using a template kindly provided by R.C.\ Dohm-Palmer. An
empirical PSF of diameter $14\arcsec$ was constructed for each image
and used to compute magnitudes for \objD. In the $I$-, $R$- and
$V$-band images the optical counterpart was detected but
unresolved. In the $B$-band image, the counterpart was not detected.

For photometric calibration we measured the flux of star \#361 from
the SA110 field described by \citet{landolt92}, within an aperture of
diameter $14\arcsec$. We adopted ``typical'' CTIO extinction
coefficients of $k_I = 0.06$, $k_R = 0.11$, $k_V = 0.15$, and $k_B =
0.28$ \citep{landolt92}. Table~\ref{tbl:photometry} lists the optical
and near-infrared total magnitudes of \objD~in 7 filters from $B$ to
$K$, using the observations described in this section and entries from
the 2MASS catalog\footnote{The Two Micron All Sky Survey (2MASS) is a
joint project of the University of Massachusetts and the Infrared
Processing and Analysis Center/California Institute of Technology,
funded by the National Aeronautics and Space Administration and the
National Science Foundation (NSF).}. Evidently \objD~is very red, with
$B-K > 10.7$. This is confirmed by the spectrophotometry of
\citet{gregg01}, who find $B-K \ga 11$.

\section{The gravitational lensing hypothesis}
\label{sec:hypothesis}

\subsection{The case for lensing}
\label{subsec:case}

The case that \objD~is gravitationally lensed relies mainly on its
near-infrared and radio morphology. Given the spectra of
\citet{gregg01}, which reveal the source to be a quasar, the
observation that the $H$-band counterpart is double
(\S\ref{subsec:nearir}) is by itself powerful evidence for lensing. At
optical and near-infrared wavelengths, quasars are almost always
observed as unresolved points. Quasars that appear double (with
separation $< 3\arcsec$) are either (1) the result of a chance
superposition between a quasar and a star, (2) a chance superposition
of quasars at different redshifts, (3) a binary quasar, or (4) a
gravitational lens.

The first possibility is ruled out by the observation that both
components are radio-loud. The second and third hypotheses are {\em a
priori} unlikely because radio-loud quasars constitute a minority
($<10\%$) of quasars generally. The observation that components A--E
all have the same continuum radio spectrum from 5~GHz to 43~GHz makes
the chance superposition hypotheses untenable, and casts serious doubt
on the binary quasar hypothesis. By contrast, the identity of spectral
indices is a natural consequence of gravitational lensing. The maximum
component separation of 681~mas is also in the range of angular sizes
that is characteristic of gravitational lensing by galaxies ($0\farcs5
- 2\arcsec$). This angular scale is set by the Einstein ring radius of
medium-redshift $L_*$ galaxies.

Furthermore, the milliarcsecond morphology of \objD~is unusual for a
GPS quasar, or indeed for any radio source. GPS sources are almost
always a single core, a compact double, a linear triple, or a core-jet
structure \citep{odea98}. Examples of VLBI observations of GPS sources
are given by \citet{snellen00}. The total angular extent is usually
less than 100 mas. By contrast, the five components of \objD~have a
much larger separation and are not collinear, which prevents any
obvious assignment of the components as cores, lobes and jets.

The curved arc between components A and B in the 1.7~GHz VLBA image
(\S~\ref{subsec:vlba}) is particularly suggestive, because the arc is
nearly perpendicular to the line from A to D. The magnification due to
gravitational lensing often results in lensed images that are
stretched tangentially, i.e., stretched along the circular or
elliptical critical curve that runs approximately through the images.
Some examples are B0218+357 \citep{patnaik95}, B1422+231
\citep{patnaik99}, MG~0414+0534 \citep{trotter00}, and B2016+112
\citep{garrett97}.

In addition, the line joining D and F is nearly perpendicular to the
line joining A and D, suggesting that D and F may also be an example
of tangential stretching. As pointed out in \S~\ref{subsec:vlba}, it
is possible that D and F are connected by an arc. In fact, if the A/B
arc is extrapolated into a full circle (to approximate a critical
curve), the circle intersects both D and F, as shown in
Figure~\ref{fig:vlba-18cm}.

Finally, Gregg et al.\ (2001) present an additional piece of
circumstantial evidence that \objD~is being gravitationally lensed:
based on its observed redshift and $K'$-band magnitude, the object
would be among the most intriniscally luminous quasars known. This
high apparent luminosity is consistent with gravitational
magnification.

\subsection{Problems with gravitational lens models}
\label{subsec:models}

An important step in the analysis of newly-discovered gravitational
lens systems is to devise a plausible model of the gravitational
potential of the foreground galaxy, and of the background source
structure, that can account for the observed image configuration. The
most common configurations by far are those with 2 and 4 lensed
images, which can often be modeled by an isothermal elliptical
potential that is lensing a single background source. However, the
radio morphology of \objD, with at least 6 components, cannot be
produced by such simple models, and we have not been able to devise a
completely satisfactory alternative model. In this section, we
describe some possible lensing scenarios, and the unsatisfactory
aspects of each one.

Is it possible that \objD~is a simple 2- or 4-image gravitational
lens, and that some of the additional radio components are actually
due to the foreground object(s) rather than the background source? Any
such scenario is unsatisfactory because A--E have the same spectral
index (\S~\ref{subsec:vla}), suggesting they are all
related. Furthermore, even if the lens galaxy is assumed to be
centered on one of the radio components, or between any two of them,
the remaining radio components are not arranged in a typical lensing
configuration.

Could all the radio components be images of a single background
source? This is unsatisfactory because lensing conserves surface
brightness, whereas two of the components (C and E) apparently have a
lower surface brightness than the other components
(\S~\ref{subsec:vlba}). One would have to invoke a propagation effect
that acts differently along the various image paths. For example,
interstellar scattering (by plasma that is either in a foreground
galaxy or our own Galaxy) might be causing differential
scatter-broadening of the images. In this scenario, the comparatively
large angular sizes of C and E, which caused them to be largely
resolved out of our VLBA images, is due to a larger column density of
electrons along those image paths. This hypothesis is testable,
because the angular sizes of the components would be expected to vary
as $\lambda^{2}$, where $\lambda$ is the observing
wavelength. Multifrequency VLBI observations that include some short
baselines ($< 10^6\lambda$) would be helpful.

The other problem with identifying all 6 components as images of a
single background source is that the image configuration cannot be
produced with simple lens models consisting of a single galaxy. Nor
have we been able to produce the image configuration with models
consisting of more than one lens galaxy, although the phase space of
parameters in such models is too large to explore comprehensively
without at least some prior constraints (e.g., the galaxy positions).

We are therefore led to consider models in which the background source
has more than one radio component.  GPS quasars commonly have more
than one radio component, so this scenario is reasonable. Because the
near-infrared counterpart is double, it is tempting to try models in
which each background source is doubly imaged. From the 1.7~GHz image
alone (Figure~\ref{fig:vlba}), such a model appears plausible. In one
scenario, the background source consists of a core and a jet with a
hot spot at its end. In one image, the core is B and the hot spot is
A; in the other, parity-reversed image, the core is D and the hot spot
is F.

A lens model consisting of a singular isothermal sphere can reproduce
the positions of A, B, D and F almost exactly. However, the predicted
magnification ratios in this model are not even close to the observed
flux ratios; the problem is that A is much brighter than F, whereas B
and D are of comparable brightness. Any lens model of this type would
need to produce a very large magnification gradient between the source
locations corresponding to A/F and B/D, which is possible for a source
almost perfectly aligned with the foreground galaxy. The main problem
with this scenario is that it ignores components C and E.

This situation, compelling evidence for gravitational lensing but
inability to model the system adequately, is frustrating but not
unique. For example, the correct model for the well-established
three-image lens B2016+112 \citep{lawrence84} has been a mystery for
over 15 years and seems to involve two lens galaxies at different
redshifts \citep{nair97}. The lensing scenario for the six-image
system B1359+154 became clear only after an image with the {\em Hubble
Space Telescope} revealed that the foreground mass was actually a
compact group of 3 galaxies \citep{rusin01}.

\section{Summary and discussion}
\label{sec:summary}

We have presented an extensive set of radio, near-IR and optical
observations of the GPS quasar \objD. The radio morphology is unusual,
with at least 5 components sharing the same radio continuum spectrum
between 5~GHz and 43~GHz. At least two of the components have $H$-band
counterparts, strongly suggesting they are lensed images of a single
source. A curved arc of radio emission between two of the components
appears to be an example of the tangential stretching that is
characteristic of gravitational lensing.

Neither the lensing correspondences between the components nor the
foreground mass distribution is clear from the present data. We
suggest three lines of observational inquiry to obtain this
information:

1.\ Multi-frequency high-resolution radio imaging.  Images at many
frequencies will allow the angular sizes of components C and E to be
measured as a function of wavelength, in order to see if they scale as
$\lambda^2$ characteristic of scatter broadening. Images at higher
frequencies will further resolve components A, B and possibly D; the
detailed morphology may be valuable in devising lens models. Finally,
sensitive images at multiple frequencies will allow the spectral index
of the dim component F to be compared to the other components.

\begin{figure}
\figurenum{5}
\plotone{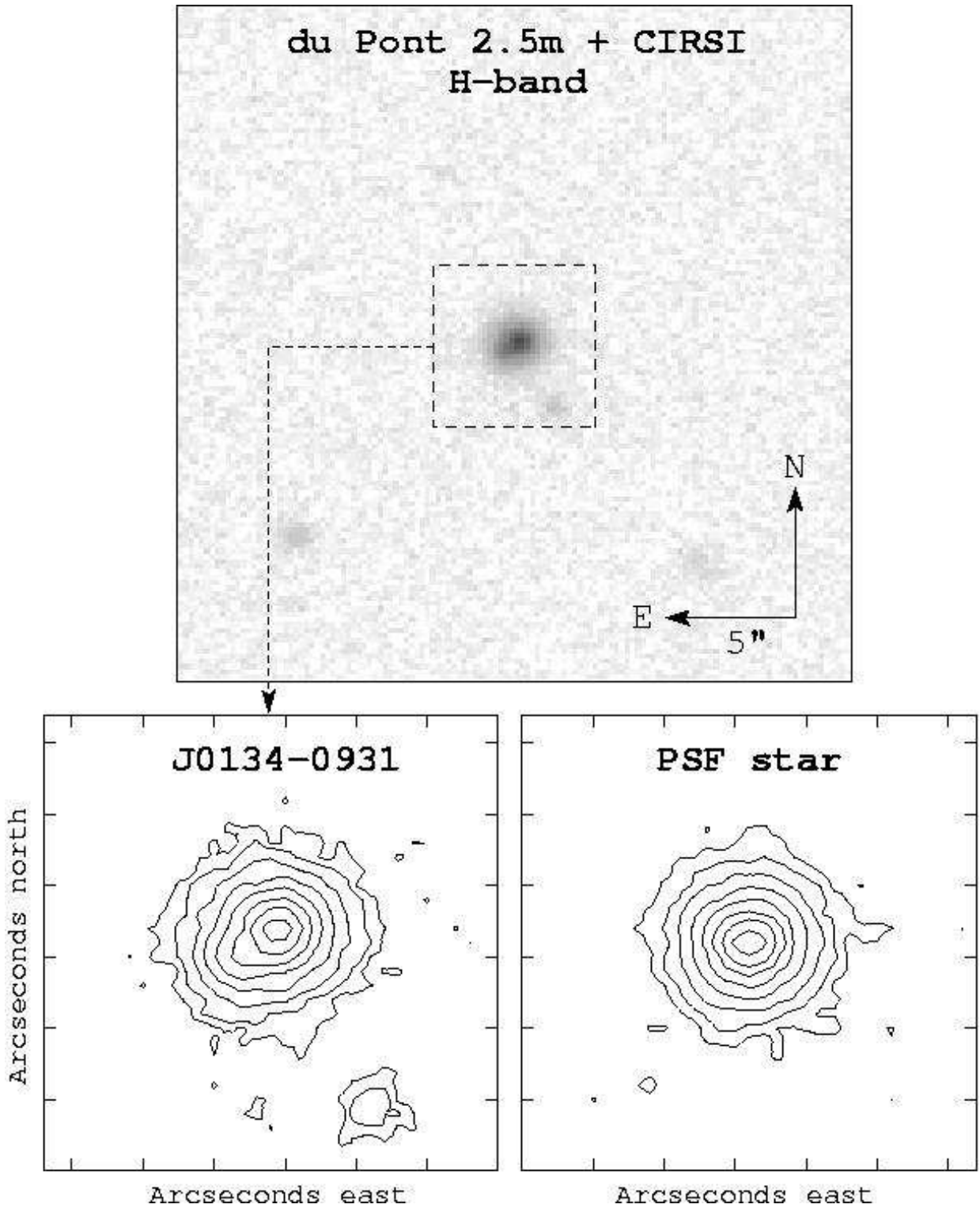}
% \vskip -0.5in
\caption{ {\bf Top.} Stacked $H$-band image ($30\arcsec\times
30\arcsec$) of \objD~(centered). {\bf Lower left.} Contour plot of the
$6\arcsec\times 6\arcsec$ region surrounding \objD. Contours begin at
$3\sigma$ and increase by factors of 2, where $\sigma$ is the RMS
noise level. {\bf Lower right.} Contour plot of the empirical PSF (see
\S~\ref{subsec:nearir}).  }
\label{fig:ir}
\end{figure}

2.\ Higher-resolution optical/near-IR imaging. This will establish
exactly which radio components have optical counterparts, and will
test for the presence of a foreground galaxy or galaxies.  Finally, it
may establish the nature of the dim component southwest of \objD~that
was detected in our near-IR image (Fig.~\ref{fig:ir}).

3.\ Optical/near-IR spectroscopy of the individual components.
Separate spectra of components A and D should verify they are both
quasars at the same redshift. Deep spectra may also reveal the
presence of foreground absorbing material, and provide its redshift.

Separating the various components at optical wavelengths will probably
require the {\em Hubble Space Telescope}. Adaptive-optics imaging is a
more challenging prospect because there are no particularly bright
stars within $30\arcsec$ to serve as a guide star.

Finally, we note that the optical counterpart of \objD~is extremely
red, with $B-K > 10.7$. This places \objD~among the recently
recognized population of red quasars that appear in radio surveys but
have been missed in optical surveys \citep[see,
e.g.,][]{francis00}. Several other gravitationally lensed radio
sources are also extremely red (e.g., MG~0414+0534, MG~J1131+0456,
JVAS~B1938+666). \citet{gregg01}, \citet{kochanek00},
\citet{webster95} and \citet{becker97}, among others, have argued that
these quasars are red due to mechanisms intrinsic to the quasar or
host galaxy rather than reddening due to foreground objects along the
line of sight, and are connected to a large population of optically
obscured AGN.

\acknowledgments

We are grateful to Bob Schommer, for obtaining optical images at Cerro
Tololo after our own attempts were spoiled by bad weather; to Tom
Muxlow and Peter Thomasson, for help with the MERLIN observations; to
Barry Clark and Steve Myers for help scheduling the VLBA observation;
and to Dick Hunstead for providing the 408~MHz measurement. We thank
Michael Gregg and collaborators for sharing their manuscript in
advance of publication. J.N.W.\ thanks the Fannie and John Hertz
foundation for supporting his graduate study, the NOAO for funding
travel to Cerro Tololo, and the Max-Planck-Institut f\"{u}r
Radioastronomie for hospitality in Bonn during part of this work.

\clearpage

\clearpage

\clearpage

\clearpage

\clearpage

\clearpage

\begin{deluxetable}{lccccccc}

\tabletypesize{\scriptsize}
\tablecaption{Radio observations of \objD\label{tbl:radio}}
\tablewidth{0pt}

\tablehead{
\colhead{Date}          &
\colhead{Observatory}   &
\colhead{Frequency}	&
\colhead{Bandwidth}	&
\colhead{Duration}	&
\colhead{Beam FWHM}	&
\colhead{RMS level}	&
\colhead{Flux scale}	\\
\colhead{}    	      	&
\colhead{}   		&
\colhead{(GHz)}		&
\colhead{(MHz)}		&
\colhead{(min)}		&
\colhead{(mas $\times$ mas, P.A.)}	&
\colhead{(mJy/beam)}	&
\colhead{uncertainty}	
}

\startdata
1992 Dec 31	&VLA	 &8.440	&100	&2	&$276\times 160$, $11\arcdeg$	&0.24 	& 3\% \\	
2000 Apr 01	&MERLIN  &4.994	&15	&36	&$128\times 39.1$, $17\arcdeg$	&1.2	& 5\% \\
2000 Apr 24	&VLBA	 &4.975	&64	&60	&$7.3\times 1.5$, $25\arcdeg$	&0.4    &5\% \\
2000 Sep 25	&ATCA	 &1.2--9.0\tablenotemark{a} &128	&2 & \nodata &\nodata &5\% \\
2000 Oct 15	&VLA	 &8.46	&100	 &10	&$297\times 195$, $23\arcdeg$	&0.15 & 3\% \\
2000 Oct 15	&VLA	 &14.94	&100	 &28	&$139\times 100$, $10\arcdeg$	&0.40 & 5\% \\
2000 Oct 15	&VLA	 &22.46	&100	 &28	&$96\times 76$, $-23\arcdeg$	&0.44 & 10\% \\
2000 Oct 15	&VLA	 &43.34	&100	 &28	&$56\times 35$, $-6\arcdeg$	&0.68 & 10\% \\
2000 Oct 31     &VLBA    &1.667  &64      &240   &$10.2\times 4.0$, $5\arcdeg$   &0.2 & 5\% \\
\enddata

\tablecomments{ The J2000 coordinates of component A of \objD~are
$01^\mathrm{h}34^\mathrm{m}35\fs667$, $-09\arcdeg31\arcmin02\farcs89$
within $0\farcs15$. }

\tablenotetext{a}{ The ATCA was used to measure total flux densities
only, with 2-minute observations at 1.216, 1.344, 1.472, 1.728, 2.240,
2.496, 4.800, 6.080, 8.640, and 9.024~GHz. }

\end{deluxetable}

\begin{deluxetable}{lccc}

\tabletypesize{\scriptsize}
\tablecaption{Radiometric model based on VLA/MERLIN data\label{tbl:radiomodels}}
\tablewidth{0pt}

\tablehead{
\colhead{Component}         &
\colhead{$\Delta$R.A.}      &
\colhead{$\Delta$Decl.}     &
\colhead{Flux ratio,}	    \\
\colhead{} &
\colhead{(mas)} &
\colhead{(mas)} &
\colhead{rel.\ to D}
}

\startdata
A & $-539.62\pm 1.20$ & $414.71\pm 1.51$ & $5.34\pm 0.13$ \\
B & $-618.80\pm 0.84$ & $264.02\pm 1.58$ & $1.317\pm 0.043$ \\
C & $-448.76\pm 1.48$ & $382.18\pm 0.69$ & $1.140\pm 0.048$ \\
D & 0 & 0 & 1 \\
E & $-263.47\pm 0.67$ & $632.79\pm 1.91$ & $0.3178\pm 0.0073$
\enddata

%\startdata
%A & 0 & 0 & $5.38\pm 0.10$ \\
%B & $-79.69\pm 0.71$ & $-151.58\pm 1.05$ & $1.30\pm 0.01$ \\
%C & $91.00\pm 0.44$  & $-32.92\pm 0.42$  & $1.16\pm 0.02$ \\
%D & $539.03\pm 0.22$ & $-415.38\pm 0.75$ & 1 \\
%E & $275.64\pm 0.85$ & $217.80\pm 1.88$  & $0.317\pm 0.008$  
%\enddata

\tablecomments{ Figures reported here are the average of the modeling
results applied to the VLA and MERLIN data. Each quoted uncertainty is
the standard deviation of these results. }

\end{deluxetable}

\begin{deluxetable}{lcccccc}

\tabletypesize{\scriptsize}
\tablecaption{Radiometric model based on 5~GHz VLBA data\label{tbl:vlba-6cm}}
\tablewidth{0pt}

\tablehead{
\colhead{Component} &
\colhead{$\Delta$R.A.} &
\colhead{$\Delta$Decl.} &
\colhead{Flux density} &
\colhead{$b_\mathrm{maj}$} &
\colhead{$b_\mathrm{min}/b_\mathrm{maj}$} &
\colhead{P.A.} \\

\colhead{} &
\colhead{(mas)} &
\colhead{(mas)} &
\colhead{(mJy)} &
\colhead{(mas)} &
\colhead{} &
\colhead{}
}

\startdata
A$_1$& $-539.93$ & $415.04$ & $269.22$ & $7.7$  & $0.28$ & $46\arcdeg$ \\
A$_2$& $-542.66$ & $409.43$ & $ 78.02$ & $20.7$ & $0.19$ & $30\arcdeg$ \\
B    & $-619.64$ & $262.95$ & $ 82.47$ & $3.7$  & $0.21$ & $20\arcdeg$ \\
C    & $-446.58$ & $382.72$ & $ 35.62$ & $11.2$ & $0.63$ & $-23\arcdeg$ \\
D    &    0      &   0      & $ 63.98$ & $2.6$  & $0.37$ & $18\arcdeg$
\enddata

\tablecomments{ Each component is an elliptical Gaussian
with the specified position, flux density, FWHM major axis ($b_\mathrm{maj})$,
ellipticity ($b_\mathrm{min}/b_\mathrm{maj}$), and position angle (degrees
east of north). }

\end{deluxetable}

\begin{deluxetable}{lcccccc}

\tabletypesize{\scriptsize}
\tablecaption{Radiometric model based on 1.7~GHz VLBA data\label{tbl:vlba-18cm}}
\tablewidth{0pt}

\tablehead{
\colhead{Component} &
\colhead{$\Delta$R.A.} &
\colhead{$\Delta$Decl.} &
\colhead{Flux density} &
\colhead{$b_\mathrm{maj}$} &
\colhead{$b_\mathrm{min}/b_\mathrm{maj}$} &
\colhead{P.A.} \\

\colhead{} &
\colhead{(mas)} &
\colhead{(mas)} &
\colhead{(mJy)} &
\colhead{(mas)} &
\colhead{} &
\colhead{}
}

\startdata
A & $-540.90$  &  $412.20$  &  $406.39$ & $51.1$ & $0.29$ & $43\arcdeg$ \\
B & $-618.83$  &  $266.58$  &  $117.48$ & $15.4$ & $0.13$ & $14\arcdeg$ \\
D &     0      &     0      &  $100.24$ & $3.8$  & $0.56$ & $28\arcdeg$ \\
F &  $-30.95$  &  $-39.14$  &  $10.82$  & $8.9$  & $0.00$ & $53\arcdeg$
\enddata

\tablecomments{ Each component is an elliptical Gaussian
with the specified position, flux density, FWHM major axis ($b_\mathrm{maj})$,
ellipticity ($b_\mathrm{min}/b_\mathrm{maj}$), and position angle (degrees
east of north). }

\end{deluxetable}

\begin{deluxetable}{lcc}

\tabletypesize{\scriptsize}
\tablecaption{Photometry of \objD\label{tbl:photometry}}
\tablewidth{0pt}

\tablehead{
\colhead{Source} &
\colhead{Filter} &
\colhead{Magnitude}
}

\startdata
CTIO 4m / Mosaic II    & B & $>24.3$ \\
CTIO 4m / Mosaic II    & V & $22.57\pm 0.07$ \\ 
CTIO 4m / Mosaic II    & R & $20.61\pm 0.05$ \\
CTIO 4m / Mosaic II    & I & $18.77\pm 0.05$ \\ 
2MASS                  & J & $16.171\pm 0.134$\\
LCO 2.5m / CIRSI       & H & $14.765\pm 0.020$\\
2MASS                  & H & $14.748\pm 0.078$\\
2MASS                  & K & $13.546\pm 0.055$
\enddata

\end{deluxetable}

\end{document}